\pdfoutput=1
\documentclass{article}

\usepackage{hyperref}

\usepackage{amssymb}
\usepackage{amsmath}
\usepackage{amsfonts}
\usepackage{upgreek}
\usepackage{graphicx,epsfig}
\usepackage{lipsum}
\usepackage{subfigure}
\usepackage{xcolor}
\usepackage{slashbox}
\usepackage{fancyhdr}
\usepackage{afterpage}
\usepackage{amsthm}
\usepackage{times,amsmath,epsfig}
\usepackage{amssymb,amsmath}
\usepackage{cite}
\usepackage{lipsum}
\usepackage{algorithmic} 
\usepackage[ruled]{algorithm} 

\begin{document}

\title{\bf Efficient Offline Waveform Design Using Quincunx/Hexagonal Time-Frequency Lattices For 5G Systems}
\author{%
{Raouia Ayadi{\small $~^{\#1}$}, In\`es Kammoun{\small $~^{*2}$}, Mohamed Siala{\small $~^{\#3}$} }%
\vspace{1.6mm}\\
\fontsize{10}{10}\selectfont\itshape
$~^{\#}$MEDIATRON Laboratory, Sup'Com, University of Carthage, 2083 Ariana, Tunisia\\
\fontsize{9}{9}\selectfont\ttfamily\upshape
$~^{1}$raouia.ayadi@supcom.tn\\
\fontsize{9}{9}\selectfont\ttfamily\upshape
$~^{3}$mohamed.siala@supcom.tn%
\vspace{1.2mm}\\
\fontsize{10}{10}\selectfont\rmfamily\itshape
$~^{*}$LETI Laboratory, ENIS, BPW 3038 Sfax, Tunisia \\
\fontsize{9}{9}\selectfont\ttfamily\upshape
$~^{2}$ines.kammoun@ieee.org
}

\maketitle

\baselineskip=14.5pt
\begin{abstract}
Conventional OFDM, adopted in LTE-A systems, cannot provide the quality of service requirements sought in 5G systems because of extreme natural channel impairments caused by higher Doppler spreads and unexpected artificial impairments caused by multi-source transmission, to be brought by 5G, and by synchronization relaxation for closed-loop signaling overhead reduction in some 5G applications. These severe  impairments induce a strong loss of orthogonality between subcarriers and OFDM symbols and, therefore, lead to a dramatic increase in ICI  and ISI. To be well armed against these dramatic impairments, we, in the present paper, optimize the transmit/receive waveforms for FBMC  systems, with hexagonal time-frequency lattices, operating over severe  doubly dispersive channels, accounting for both natural and artificial impairments. For this, we exploit the POPS paradigm, recently proposed for rectangular time-frequency lattices, to design offline waveforms maximizing the SINR for hexagonal time-frequency lattices. We show that FBMC, with  hexagonal lattices, offers a strong improvement in SIR with respect to conventional OFDM and an improvement of $1\hspace{0.1cm}\mbox{dB}$  with respect to POPS-FBMC, with classical rectangular lattices. Furthermore, we show that the hexagonal POPS-FBMC brings more robustness to frequency synchronization errors and offers a $10\hspace{0.1cm}\mbox{dB}$ reduction in OOB emissions with respect to rectangular POPS-FBMC.
\end{abstract}

\section{Introduction}
Orthogonal frequency division multiplexing (OFDM) is now a well established technique that provides high-data-rate wireless communications, through a transformation of the frequency-selective channel into several non-selective
sub-channels, thereby reducing the inter-symbol interference (ISI)~\cite{Saltzberg}. This modulation has received a great success in several standards, such as asymmetric digital subscriber line (ADSL), IEEE 802.11a/g (WiFi), IEEE
802.16 (WiMAX), digital audio broadcasting (DAB), digital video
broadcasting (DVB) and long term evolution-advanced (LTE-A)~\cite{hara}. For all these systems, OFDM uses a rectangular waveform, in order to maintain maximum spectrum efficiency
and ensure orthogonality between the different shifts of the used waveform in the time-frequency plane. Unfortunately, in practice, the mobile radio channel is very dispersive in time and frequency, causing an orthogonality loss between the time-frequency shifted versions of this rectangular waveform. By adding
a cyclic prefix (CP), the conventional OFDM system
becomes less sensitive to delay spread and time synchronization errors, at the cost of a spectrum
efficiency loss. However, in the presence of Doppler spread and
frequency synchronization errors, the bad frequency localization of the rectangular waveform leads to a very important inter-carrier interference (ICI), in both downlink and uplink channels. Moreover, at the uplink channel, guard bands between users accessing the spectrum resources are required because of the asynchronous nature of multiple access and the strong out-of-band (OOB) power leakages between adjacent user bands, leading to an inefficient use of spectrum resources. In addition, conventional OFDM has shown its limitations in some specific, yet important, applications, such as LTE-A coordinated multi-point (CoMP)~\cite{Wild} and multimedia broadcast multicast services (MBMS), since it requires the use of an extended CP and incurs a significant reduction in spectrum efficiency.\\
With respect to LTE-A, the future 5G wireless cellular system is expected to support new applications and services, such as tactile internet, the internet of things (IoT) and machine type communications (MTC). Moreover, new constraints and objectives, different from those of LTE-A, have been defined for 5G, such as 1000 times higher data rates, 100 times higher number of connected devices, 10 times longer battery life and 5 times reduced latency~\cite{GFettweis1, Wunder1, Wunder}. To address all these requirements, it is necessary to alleviate the synchronization mechanism overhead for small packet transmissions, brought by tactile internet, IoT and MTC, in order to reduce latency and efficiently use energy and radio spectrum resources. Unfortunately, diminishing signaling by synchronization reduction or relaxation introduces artificial impairments, caused by timing and carrier frequency offset errors. In addition, other extra artificial impairments, caused by multi-source transmissions,
such as a CoMP, MBMS and
cloud-radio access networks (C-RAN), are expected to be prevalent and common. Besides, 5G has to be immune to natural impairments caused by larger Doppler spreads brought by the expected
use of higher frequency millimeter-wave (mm-wave) bandwidths. These delay and frequency spreads can't be tolerated and endured by conventional OFDM, leading to either a dramatic loss in spectrum efficiency or a strong decrease in signal-to-interference-and-noise ratio (SINR). \\
To face all previously mentioned impairments, a plethora of research activities have focused on waveform design for 5G with either waveform optimization~\cite{Ayadi,Ayadi2}, flexible waveform configuration~\cite{Jiay} or application of lapped transform \cite{Bellanger}. In~\cite{Ayadi}, based on SINR maximization, the authors proposed a novel approach to design continuous-time transmit/receive waveforms. The proposed waveforms, which are expressed as linear combinations of
the most time-frequency localized Hermite waveforms, demonstrated an extra robustness to doubly dispersive channels. To reduce the complexity of their optimization, the same authors proposed, in~\cite{Ayadi2}, an approximation of the SINR, using Taylor's series, truncated at the
second order, of the interference and the useful signal mean powers. However, the complexity of the continuous-time optimization is still important and the convergence to the global maximum is not only arduous but also not guaranteed. In~\cite{Jiay}, an enabler for flexible waveform use, called filtered OFDM (f-OFDM), has been proposed, based on a flexible arrangement of the time-frequency plane and the use of different waveforms in different subbands, depending on services requirements. This technique has many advantages over OFDM, ranging from relaxed synchronization and support of inter-subband asynchronous transmission to reduced OOB emissions. In~\cite{Bellanger}, the authors proposed a new multicarrier scheme based on the lapped transform. While mainting some desired characteristics of conventional OFDM, this scheme responds to some critical objectives for 5G, such as spectral efficiency increase, asynchronous transmission support and spectrum protection required by users in cognitive radio scenarios. Furthermore, some waveforms suitable for 5G applications, such as filter bank multicarrier (FBMC), generalized frequency division multiplexing (GFDM) and universal filtered multicarrier (UFMC), have been proposed in recent projects, such as PHYDYAS~\cite{Bellenger} and 5GNOW~\cite{Wunder}. In FBMC, the subcarriers are filtered with a well time and frequency localized waveform in order to reduce OOB emissions, without the requirement to add a CP~\cite{Cassiau}. Compared to conventional OFDM, FBMC consumes less radio resources, for the same SINR, and, as a consequence, offers an increased immunity to interferences, and hence a higher SINR, for the same spectrum efficiency. However, FBMC isn't suitable for low latency scenarii and requires a higher processing complexity, ranging from digital filtering and channel estimation to time and frequency synchronization, in order to reduce ISI and ICI. As another alternative, GFDM has been proposed as a generalization of OFDM, by arranging the data in a two-dimensional time-frequency block
structure, using a flexible waveform and adding a short CP for an entire block of OFDM symbols~\cite{Fettweis}. GFDM offers a better control on OOB emissions, relative to conventional OFDM, and offers a suitable and friendly use of fragmented spectrum. Unfortunately, it is not robust in the presence of Doppler spread and frequency synchronization errors, especially in the presence of long blocks. In the third proposed multicarrier modulation technique UFMC, the pulse shaping is applied to a group of contiguous subcarriers, to increase system robustness to synchronization errors and to better support fragmented spectrum~\cite{Vakilian}. Regrettably, the required extra filtering generally inflicts severe attenuations to the edge subcarriers with respect to the middle ones. Moreover, it comes at the price of a reduced tolerance to multipath delay spread, for a given guard interval time overhead.  \\
This paper is concerned with the optimization of FBMC waveforms for hexagonal time-frequency lattices. Indeed, several studies in the scientific literature have announced that the communication quality on hexagonal time-frequency lattices outperforms that on conventional rectangular lattices~\cite{Strohmer,Han2,Han}. The proposed optimization method corresponds to an extension of the method, proposed in~\cite{siala} by Siala \textit{et al.}, on rectangular time-frequency lattices, to hexagonal lattices, in order to have an additional immunity to interferences. Compared to the optimization method proposed in~\cite{Ayadi} and~\cite{Ayadi2}, the strategy adopted in this paper for waveform optimization offers more simplicity and accuracy and provides discrete-time waveforms that are directly implementable in hardware. \\
Performed offline, the proposed iterative optimization strategy can be carried for a finite bunch of representative propagation channel dispersion statistics, for given all expected 5G services (such as CoMP, MBMS, C-RAN, $\cdots$). The resulting optimized transmit-receive waveform pairs form a codebook and are then used online, in an adaptive fashion, to better adapt to the slowly varying channel propagation statistics. More specifically, as in adaptive modulation and coding (AMC), adaptive waveform communications (AWC) could become a reality in 5G, whereby the most suitable pair of transmit-receive waveforms is selected from the codebook and used, as a function of an estimate of the current channel dispersion statistics.\\
The rest of this paper is organized as follows. In Section II, we introduce the system model. We analyse the interference and
noise statistical characteristics in Section III. In Section IV, we present the optimization procedure of the transmit/receive waveforms based on SINR maximization. Then, in Section V, we detail the approach for signal and interference Kernels computation in Section V. Finally, we devote Section VI to simulation results and Section VII to conclude this paper. \\
Throughout the paper, the norms of vectors are denoted
by $\left\|.\right\|$ and the Hermitian scalar product of two vectors is represented by $\left\langle .,.\right\rangle$. The operators $(.)^*$, $(.)^{H}$, $(.)^T$
and $\mathbb E(.)$ stand for
transposition, complex conjugation, transconjugation and expectation, respectively. We denote by ${\otimes}$ the  Kronecker matrix product and by $\odot$ the component-wise product (a.k.a. the Hadamard matrix product). The notations, $\left(.\right)_q$ and $\left(.\right)_{pq}$, are used to refer to vector and matrix with components and entries generically indexed by $q$ and $(p,q)$, respectively. The function, $[\textbf W,\boldsymbol{\Lambda}]=\mbox{eig}(\textbf A)$, produces a diagonal matrix $\boldsymbol{\Lambda}$ of eigenvalues and a matrix $\textbf W$ whose columns are the corresponding eigenvectors of matrix $\textbf A$, while the function, $[\textbf w,\lambda]=\mbox{eigs}(\textbf A)$, returns the eigenvector $\textbf w$ associated to the maximum eigenvalue $\lambda$ of matrix $\textbf A$. Finally, $\textbf{I}_m$ represents the $m{\times}m$ identity matrix, with ones in the diagonal and zeros elsewhere, and $J_0(.)$ denotes the $0^{\mathrm{th}}$-order Bessel function of the first kind.

\section{System model}
\subsection{Transmitter and receiver models}
\begin{figure}[H]
    \centering
\includegraphics[width=0.51\textwidth]{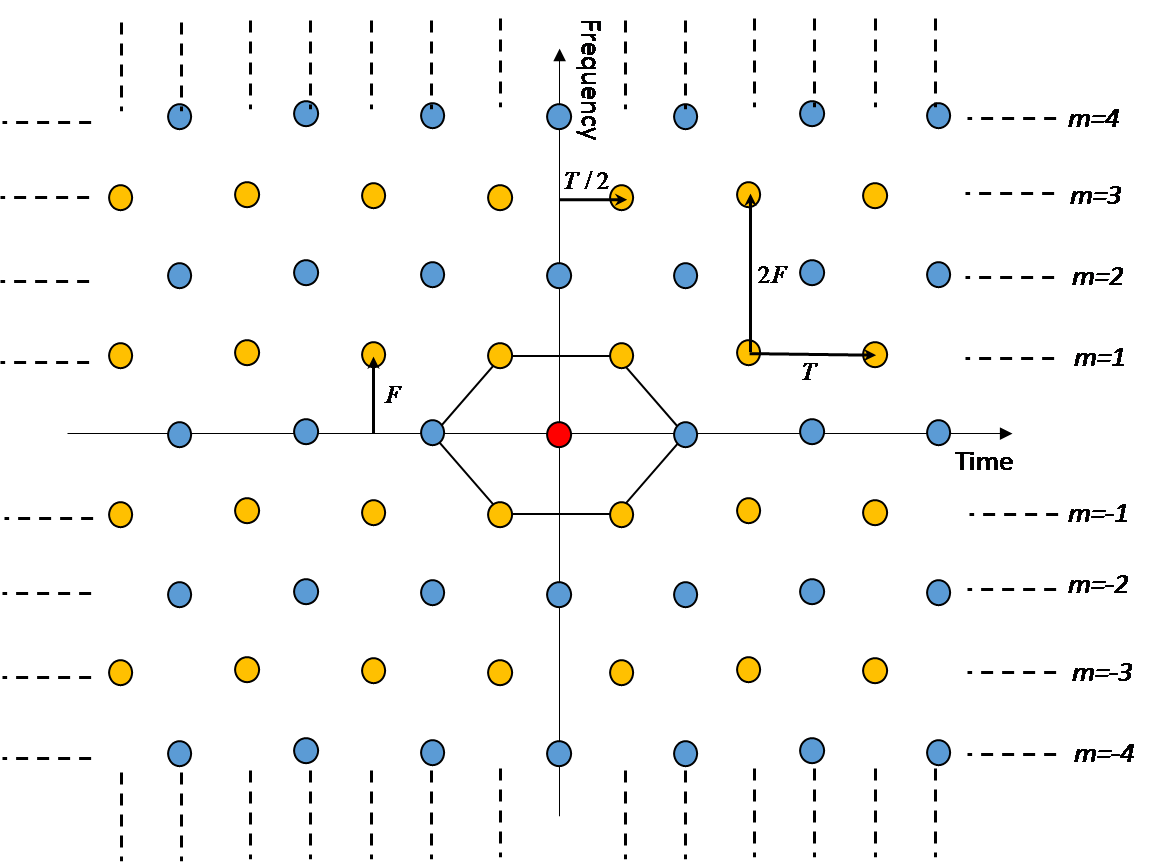}
    \caption{Hexagonal lattices in the time-frequency plane.}
    \label{schemahexagonal}
\end{figure}

We consider a baseband model of a multicarrier
system  with  $Q$  subcarriers, regularly  spaced  by  $F$  in  frequency. The transmitted multicarrier signal is sampled at a sampling rate $R_s=1/T_s$, where $T_s=T/N$ is the sampling period, $T$ is the OFDM symbol period and $N$ is an integer accounting for the number of samples per OFDM symbol period. Due to the hexagonal nature of the time-frequency lattice and the underlying half-symbol period shift between consecutive subcarriers, $N$ must be even, leading to a slight flexibility reduction with respect to the rectangular time-frequency lattice, where $N$ could also be odd. The subcarrier spacing, $F$, is related to the sampling period and to the number of subcarriers by $1/F=QT_s$. In this study, the time-frequency lattice density, $\Delta=1/FT$, is taken below unity, leading to an undersampled time-frequency lattice in the Weyl-Heisenberg frame theory jargon. The undersampled nature of the lattice, acquired by taking $Q$ smaller than $N$, offers flexibility and room to absorb and put up with different impairments inflicted by the channel and the transmitter and receiver imperfections. It is also to be noted that the positive difference $(N-Q)T_s$ is equivalent to the notion of guard interval time in conventional OFDM. \\
Working in the discrete-time domain at both transmitter and receiver, the sampled version of the transmitted signal is represented by the infinite vector
\begin{equation}\label{eq1}
\textbf{e}=\left(\cdots, e_{-2}, e_{-1}, e_{0}, e_{1}, e_{2},\cdots\right)^T=\left(e_{q}\right)_{q{\in}\mathbb{Z}}^T,
\end{equation}
where $e_{q}$ is the transmitted signal sample at time $qT_s$, with $q\in\mathbb{Z}$. This vector can be written as
\begin{equation}\label{1}
\textbf{e}=\displaystyle{\sum_{m,n}}a_{mn}\boldsymbol{\upvarphi}_{mn},
\end{equation}
where the function, $\boldsymbol{\upvarphi}_{mn}$, used for the transmission of symbol $a_{mn}$, results from a time shift, by $t_{mn}$, and a frequency shift, by $f_{mn}$, of the transmit waveform vector $\boldsymbol{\upvarphi}$, and $a_{mn}$, $m, n \in \mathbb{Z}$, are i.i.d. modulated symbols, of zero mean and common mean
transmit energy $E_s=\mathbb{E}\left[\left|a_{mn}\right|^2\right]\left\|\boldsymbol{\upvarphi}\right\|^2$. As illustrated in Fig.~\ref{schemahexagonal}, the time and frequency shifts, dictated by quincunx/hexagonal lattices, are determined by the generator matrix
\begin{equation}
\left(\begin{array}{c}
   t_{mn}
   \cr f_{mn}
\end{array}\right)= \left(\begin{array}{ccc}
   T & T/2\\
   \cr 0& F
\end{array}\right)\left(\begin{array}{c}
   n
   \cr m
\end{array}\right).
\end{equation}

Consequently, the shifted waveforms are defined by
	\begin{equation} \begin{array}{lllllllll}
\boldsymbol{\upvarphi}_{mn}=\boldsymbol{\upsigma}_{(n+\frac{m}{2})N}(\boldsymbol{\upvarphi}){\odot}\left(e^{j2{\pi}mq/Q}\right)_q,
\end{array}
\end{equation}
where $\boldsymbol{\upsigma}_{k}(\boldsymbol{\upvarphi})$ is obtained by shifting $\boldsymbol{\upvarphi}$ by $k$ samples, corresponding to a time shift by $kT_s$. In order to reduce the optimization complexity and the latency, the transmit waveform is assumed to have a finite duration, $D_{\boldsymbol{\upvarphi}}$, and thus a finite number of samples, $N_{\boldsymbol{\upvarphi}}=D_{\boldsymbol{\upvarphi}}/T_s$, where $N_{\boldsymbol{\upvarphi}}$ is a positive integer.\\	
By considering a time-varying channel, the sampled version of the received signal, $\textbf r=(r_q)_{q{\in}\mathbb{Z}}$, is determined by

\begin{equation}\label{2}
r_q=\displaystyle{\sum_{p}}h(p,q)e_{q-p}+n_{q}=\displaystyle{\sum_{mn}}a_{mn}\left(\tilde{\boldsymbol{\upvarphi}}_{mn}\right)_{q}+n_{q},
\end{equation}
where $\left(\tilde{\boldsymbol{\upvarphi}}_{mn}\right)_{q}=\sum_{p}h(p,q)\left(\boldsymbol{\upvarphi}_{mn}\right)_{q-p}$ is the $q^{\mathrm{th}}$ sample of the channel-distorted version, $\tilde{\boldsymbol{\upvarphi}}_{mn}$, of $\boldsymbol{\upvarphi}_{mn}$, $h(p,q)$ is the impulse response of the discrete channel at time
$qT_s$ and $n_q$ is a  discrete-time complex  additive  white  Gaussian  noise (AWGN), the samples of which are centered, uncorrelated, with common variance $N_0$.\\
The decision variable on the transmitted symbol $a_{mn}$ is obtained by
\begin{equation}\label{decision-symbol}
\Lambda_{mn}= \left\langle \boldsymbol{\uppsi}_{mn},\textbf r\right\rangle=\boldsymbol{\uppsi}_{mn}^{H}\textbf r,
\end{equation}
where $\boldsymbol{\uppsi}_{mn}$ is a time-frequency shifted version of the receive waveform, $\boldsymbol{\uppsi}$, defined by
\begin{equation} \begin{array}{lllllllll}
\boldsymbol{\uppsi}_{mn}=\boldsymbol{\upsigma}_{(n+\frac{m}{2})N}(\boldsymbol{\uppsi}){\odot}\left(e^{j2{\pi}mq/Q}\right)_q.
\end{array}
\end{equation}
Likewise, the receive waveform is assumed to have a finite duration, $D_{\boldsymbol{\uppsi}}$, and thus a finite number of samples, $N_{\boldsymbol{\uppsi}}=D_{\boldsymbol{\uppsi}}/T_s$, which is taken as a positive integer. In order to better adapt to the possible discrepancy in the complexity capabilities of both transmitter and receiver, we can consider different durations of the transmit and receive waveforms.
\subsection{Channel model}
The discrete-time channel is assumed to satisfy the Wide-Sense Stationary Uncorrelated
Scattering (WSSUS) criterion, with discrete autocorrelation function~\cite{Proakis}
\begin{equation}\label{3}
\phi_h(p_1,p_2;{\Delta}q)=
E\left[h^*(p_1;q)h(p_2;q+{\Delta}q)\right]=\phi_h(p_1;{\Delta}q)\delta_K(p_2-p_1),
\end{equation}
where $\delta_K(.)$ is the Kronecker symbol, $\phi_h(p)=\phi_h(p;0)$ is the channel multipath power profile and

\begin{equation}\label{4}S(p,{\nu})=
\displaystyle{\sum_{{\Delta}q}}\phi_h(p,{\Delta}q)e^{-2j\pi{\nu}T_s{\Delta}q}
\end{equation}
is the channel scattering function. \\
For a simplified derivation of the expression of the SINR, to be maximized as a function of the transmit and receive waveforms, we start by considering  a channel with a finite number, $K$, of paths, with channel impulse response
\begin{equation}\label{5}h(p,q)=\displaystyle{\sum_{k=0}^{K-1}}h_k\exp(j2{\pi}\nu_kT_sq)\delta_K(p-p_k),
\end{equation}
where $h_k$, $\nu_k$ and $p_k$ are respectively the amplitude, the Doppler frequency shift and the normalized time delay shift of the
$k^{\mathrm{th}}$ path, with a time delay $p_kT_s$. The paths amplitudes, $h_k$, are assumed to be centered and decorrelated random complex Gaussian variables with average powers $\pi_k=\mathbb{E}\left[\left|h_k\right|^2\right]$. The channel scattering function is therefore given by
\begin{equation}\label{6}S(p,{\nu})=\displaystyle{\sum_{k=0}^{K-1}}\pi_k\delta_K(p-p_k)\delta({\nu}-{\nu}_k).
\end{equation}
Notice that this function can take into account the cumulative effects of natural channel impairments, due to the delay and frequency spreads, and artificial impairments, due to the time and frequency synchronization errors and asynchronism (a.k.a. misalignment) between users in the uplink. \\
Without any loss of generality, we assume the channel to be of normalized average power, meaning that $\sum_{k=0}^{K-1}\pi_k=1$, in order to maintain the same average energy per symbol at the transmission and reception sides.
\section{Analysis of average useful, interference and noise powers}
For simplicity sake, and without loss of generality, we focus on the reception of symbol
$a_{00}$. While referring to~(\ref{decision-symbol}), $\Lambda_{00}$ is expressed as
\begin{equation}\label{8}\begin{array}{llll}
\Lambda_{00}&=\langle \boldsymbol{\uppsi}_{00},\displaystyle{\sum_{mn}{a}_{mn}}\tilde{\boldsymbol{\upvarphi}}_{mn}\rangle
+ \left\langle \boldsymbol{\uppsi}_{00},\textbf{n}\right\rangle \\
&={a}_{00}\left\langle \boldsymbol{\uppsi}_{00},\tilde{\boldsymbol{\upvarphi}}_{00}\right\rangle+\displaystyle{\sum_{(m,n)\neq(0,0)}}{a}_{mn}\left\langle \boldsymbol{\uppsi}_{00},\tilde{\boldsymbol{\upvarphi}}_{mn}\right\rangle+
\left\langle \boldsymbol{\uppsi}_{00},\textbf{n}\right\rangle,
\end{array}
\end{equation}
where $\textbf n=(n_q)_{q{\in}\mathbb{Z}}$ is the received noise vector. We note that the decision variable $\Lambda_{00}$ is composed of
a desired term as well as interference and noise terms. Based on this expression of the decision variable, we can define the useful signal power as
\begin{equation}\label{power-signal}
{P_S}=\frac{E_s}{\left\|\boldsymbol{\upvarphi}\right\|^2}\mathbb{E}\left[\left|\left\langle \boldsymbol{\uppsi}_{00},\tilde{\boldsymbol{\upvarphi}}_{00}\right\rangle\right|^2\right] =\frac{E_s}{\left\|\boldsymbol{\upvarphi}\right\|^2}\boldsymbol{\uppsi}_{00}^H\mathbb{E}\left[\tilde{\boldsymbol{\upvarphi}}_{00}\tilde{\boldsymbol{\upvarphi}}^H_{00}\right]\boldsymbol{\uppsi}_{00}=\frac{E_s}{\left\|\boldsymbol{\upvarphi}\right\|^2}\boldsymbol{\uppsi}^H\mathrm{\textbf{KS}}_{S(p,\nu)}^{\boldsymbol{\upvarphi}}\boldsymbol{\uppsi},
\end{equation}
with $\mathrm{\textbf{KS}}_{S(p,\nu)}^{\boldsymbol{\upvarphi}}$ being the useful Kernel matrix defined by
\begin{equation}\label{useful-kernel}\mathrm{\textbf{KS}}_{S(p,\nu)}^{\boldsymbol{\upvarphi}}=\displaystyle{\sum_{k=0}^{K-1}}\pi_k\boldsymbol{\upphi}_{\nu_k}\odot
\left(\boldsymbol{\upsigma}_{p_k}(\boldsymbol{\upvarphi})\boldsymbol{\upsigma}_{p_k}(\boldsymbol{\upvarphi})^H\right),
\end{equation}
where $\boldsymbol{\upphi}_{\nu_k}$ is the Hermitian matrix with $(p,q)^{\mathrm{th}}$ entry $e^{2j{\pi}T_s\nu_k(p-q)}$. Similarly, by considering the decorrelated and centered nature of the transmitted symbols, we can write the mean power of the interference term as
\begin{equation}\label{power-interference}
\begin{array}{lll}
{P_I}&=\frac{E_s}{\left\|\boldsymbol{\upvarphi}\right\|^2}\displaystyle{\sum_{(m,n)\neq(0,0)}}\mathbb{E}\left[\left|\left\langle \boldsymbol{\uppsi}_{00},\tilde{\boldsymbol{\upvarphi}}_{mn}\right\rangle\right|^2\right]\\ &=\frac{E_s}{\left\|\boldsymbol{\upvarphi}\right\|^2}\displaystyle{\sum_{(m,n)\neq(0,0)}}\boldsymbol{\uppsi}_{00}^H\mathbb{E}\left[\tilde{\boldsymbol{\upvarphi}}_{mn}\tilde{\boldsymbol{\upvarphi}}^H_{mn}\right]\boldsymbol{\uppsi}_{00}&=\frac{E_s}{\left\|\boldsymbol{\upvarphi}\right\|^2}\boldsymbol{\uppsi}^H\mathrm{\textbf{KI}}_{S(p,\nu)}^{\boldsymbol{\upvarphi}}\boldsymbol{\uppsi},
\end{array}
\end{equation}
where $\mathrm{\textbf{KI}}_{S(p,\nu)}^{\boldsymbol{\upvarphi}}$ is the inter-symbol interference Kernel, given by
\begin{equation}\label{interference-kernel}
\mathrm{\textbf{KI}}_{S(p,\nu)}^{\boldsymbol{\upvarphi}}=\displaystyle{\sum_{k=0}^{K-1}}\pi_k\boldsymbol{\upphi}_{\nu_k}\odot\displaystyle{\sum_{(m,n)\neq(0,0)}}
\left(\boldsymbol{\upsigma}_{p_k}(\boldsymbol{\upvarphi}_{mn})\boldsymbol{\upsigma}_{p_k}(\boldsymbol{\upvarphi}_{mn})^H\right).
\end{equation}
Finally, the average power of the noise term is expressed as
\begin{equation}\label{power-noise}
{P_N}=\mathbb{E}\left[\left|\left\langle \boldsymbol{\uppsi}_{00},\textbf n\right\rangle\right|^2\right]=\boldsymbol{\uppsi}_{00}^H\mathbb{E}\left[\textbf{n}\textbf{n}^H\right]\boldsymbol{\uppsi}_{00}=N_0{\left\|\boldsymbol{\uppsi}\right\|^2}.
\end{equation}
We note that both Kernel matrices, $\mathrm{\textbf{KS}}_{S(p,\nu)}^{\boldsymbol{\upvarphi}}$ and $\mathrm{\textbf{KI}}_{S(p,\nu)}^{\boldsymbol{\upvarphi}}$, are Hermitian positive-definite. To derive the POPS optimization algorithm, to be presented in Section~\ref{sect-opt}, we also note that these Kernels obey the equalities
\begin{equation}\label{proof1}
\boldsymbol{\uppsi}^H\mathrm{\textbf{KS}}_{S(p,\nu)}^{\boldsymbol{\upvarphi}}\boldsymbol{\uppsi}=\boldsymbol{\upvarphi}^H\mathrm{\textbf{KS}}_{ S(-p,-\nu)}^{\boldsymbol{\uppsi}}\boldsymbol{\upvarphi}
\end{equation}
and
\begin{equation}\label{proof2}
\boldsymbol{\uppsi}^H\mathrm{\textbf{KI}}_{S(p,\nu)}^{\boldsymbol{\upvarphi}}\boldsymbol{\uppsi}=\boldsymbol{\upvarphi}^H\mathrm{\textbf{KI}}_{S(-p,-\nu)}^{\boldsymbol{\uppsi}}\boldsymbol{\upvarphi}.
\end{equation}
Consequently, given the receiver waveform $\boldsymbol{\uppsi}$, we can write the useful signal power and  the mean power of the interference as quadratic functions in the transmit waveform $\boldsymbol{\upvarphi}$, with the new Kernels $\mathrm{\textbf{KS}}_{S(-p,-\nu)}^{\boldsymbol{\uppsi}}$ and $\mathrm{\textbf{KI}}_{S(-p,-\nu)}^{\boldsymbol{\uppsi}}$, respectively. Again, using these new Kernels, we can optimize the transmitter waveform $\boldsymbol{\upvarphi}$ through a maximization of the SINR.
\section{Optimization procedure}\label{sect-opt}
The SINR, defined as the ratio of the mean power of the useful signal, in~(\ref{power-signal}), and the mean power of the interference and noise terms, in~(\ref{power-interference}) and~(\ref{power-noise}), respectively, is used as the optimization criterion. The maximization of this SINR leads to the optimum pair of transmit and receive waveforms
\begin{equation}\label{opt-prb}
\left(\boldsymbol{\upvarphi}^{opt},\boldsymbol{\uppsi}^{opt}\right)={\arg}\hspace{0.1cm}{\max_{\boldsymbol{\upvarphi},\boldsymbol{\uppsi}}}\hspace{0.1cm}\mathrm{SINR}.
\end{equation}
To determine the couple $\left(\boldsymbol{\upvarphi},\boldsymbol{\uppsi}\right)$ that maximizes (\ref{opt-prb}), we express explicitly the SINR in function of $\boldsymbol{\upvarphi}$ and $\boldsymbol{\uppsi}$ as follows
\begin{equation}
\begin{array}{lll}
\mathrm{SINR}=\frac{P_S}{P_I+P_N}=\frac{\boldsymbol{\uppsi}^H\mathrm{\textbf{KS}}_{S(p,\nu)}^{\boldsymbol{\upvarphi}}\boldsymbol{\uppsi}}
{\boldsymbol{\uppsi}^H\mathrm{\textbf{KI}}_{S(p,\nu)}^{\boldsymbol{\upvarphi}}\boldsymbol{\uppsi}+\mathrm{SNR}^{-1}\left\|\boldsymbol{\upvarphi}\right\|^2\left\|\boldsymbol{\uppsi}\right\|^2}=\frac{\boldsymbol{\upvarphi}^H\mathrm{\textbf{KS}}_{S(-p,-\nu)}^{\boldsymbol{\uppsi}}\boldsymbol{\upvarphi}}
{\boldsymbol{\upvarphi}^H\mathrm{\textbf{KI}}_{S(-p,-\nu)}^{\boldsymbol{\uppsi}}\boldsymbol{\upvarphi}+\mathrm{SNR}^{-1}\left\|\boldsymbol{\uppsi}\right\|^2\left\|\boldsymbol{\upvarphi}\right\|^2},
\end{array}
\end{equation}
where $\mathrm{SNR}=E_s/N_0$ is the signal to noise ratio. In some favorable transmission scenarii, the noise term is negligible with respect to the interference term, resulting in an measure of the SIR, as a substitute of the SINR. Let

\begin{equation}\label{eq_kin}
\mathrm{\textbf{KIN}}_{S(p,\nu)}^{\boldsymbol{\upvarphi}}=\mathrm{\textbf{KI}}_{S(p,\nu)}^{\boldsymbol{\upvarphi}}+\mathrm{SNR}^{-1}\left\|\boldsymbol{\upvarphi}\right\|^2{\textbf I}_{N_{\boldsymbol{\uppsi}}}
\end{equation}
be the interference plus noise Kernel used in the quadratic form of the SINR denominator. Referring to (\ref{proof1}) and (\ref{proof2}), the optimization problem in~(\ref{opt-prb}) is equivalent to a maximization of a generalized Rayleigh quotient involving an optimization of the receive waveform $\boldsymbol{\uppsi}$, for a given transmit waveform $\boldsymbol{\upvarphi}$, and an optimization of the transmit transmit waveform
$\boldsymbol{\upvarphi}$, for a given receive waveform $\boldsymbol{\uppsi}$. Several algorithms can be used for an iterative offline optimization of this problem. Thanks to its proven numerical stability, the SINR optimization algorithm adopted in this paper consists in diagonalizing the denominator $\mathrm{\textbf{KIN}}_{S(p,\nu)}^{\boldsymbol{\upvarphi}}$ and making a basis change, so that the maximization amounts to finding the maximum-eigenvalue eigenvector of a quadratic form. More precisely, we first decompose the Kernel $\mathrm{\textbf{KIN}}_{S(p,\nu)}^{\boldsymbol{\upvarphi}}$ as

\begin{equation}\label{4}
\mathrm{\textbf{KIN}}_{S(p,\nu)}^{\boldsymbol{\upvarphi}}=\textbf{U}\boldsymbol{\Lambda}\textbf{U}^H,
\end{equation}
where $\textbf{U}$ is a unitary matrix and $\boldsymbol{\Lambda}$ is a diagonal matrix with non-negative
eigenvalues corresponding to the eigenvectors of $\mathrm{\textbf{KIN}}_{S(p,\nu)}^{\boldsymbol{\upvarphi}}$. Second, we make a basis change in the Kernel $\mathrm{\textbf{KIN}}_{S(p,\nu)}^{\boldsymbol{\upvarphi}}$ as

\begin{equation}\label{4}
\boldsymbol{\uppsi}^H\mathrm{\textbf{KIN}}_{S(p,\nu)}^{\boldsymbol{\upvarphi}}\boldsymbol{\uppsi}=\boldsymbol{\uppsi}^H\textbf{U}\boldsymbol{\Lambda}\textbf{U}^H\boldsymbol{\uppsi}=\textbf{u}^H\textbf{u},
\end{equation}
with the introduction of the vector $\textbf{u}=\boldsymbol{\Lambda}^{\frac{1}{2}}\textbf{U}^H\boldsymbol{\uppsi}$.\\
Banking on the fact that $\mathrm{\textbf{KI}}_{S(p,\nu)}^{\boldsymbol{\upvarphi}}$ is a positive Hermitian matrix, while referring to~(\ref{eq_kin}), we are sure that all eigenvalues of $\boldsymbol{\Lambda}$ are greater than or equal to $\mathrm{SNR}^{-1}$ and therefore are strictly positive. Hence, we can recover $\boldsymbol{\uppsi}$ from $\textbf{u}$ as $\boldsymbol{\uppsi}=\frac{\textbf{U}\boldsymbol{\Lambda}^{-\frac{1}{2}}\textbf{u}}{\|\textbf{U}\boldsymbol{\Lambda}^{-\frac{1}{2}}\textbf{u}\|}$, up to a normalizing multiplicative factor, avoiding any numerical instability. By replacing this new expression of the receive waveform $\boldsymbol{\uppsi}$ in the expression of the SINR, we end up with
\begin{equation}\label{4}
\mathrm{SINR}=\frac{\textbf{u}^H\boldsymbol{\Phi}\textbf{u}}{\textbf{u}^H\textbf{u}},
\end{equation}
with $\boldsymbol{\Phi}=\boldsymbol{\Lambda}^{-\frac{1}{2}}\textbf{U}^H\mathrm{\textbf{KS}}_{S(p,{\nu})}^{{\boldsymbol{\upvarphi}}^{(k)}}\textbf{U}\boldsymbol{\Lambda}^{-\frac{1}{2}}$ being a positive Hermitian matrix. Consequently, the maximization of the SINR consists in finding the maximum eigenvalue of $\boldsymbol{\Phi}$ and its associated eigenvector $\textbf{u}$. The stages of this optimization approach are detailed in Algorithm~\ref{secondapproch}.\\
In perfect agreement with the POPS paradigm introduced in~\cite{siala}, Algorithm~\ref{secondapproch} is an offline iterative algorithm, composed of an initialization stage and an iterative stage. In the initialization stage, the good choice of the transmit waveform initialization is very critical in order to ensure the convergence of the algorithm to a global maximum of the SINR, that results in the best optimized transmit and receive waveforms. In each iteration $k$ of the iterative stage, we proceed through two major steps:
\begin{itemize}
\item A first step dedicated to the optimization of the receive waveform, $\boldsymbol{\uppsi}^{(k)}$, given the previously obtained transmit waveform $\boldsymbol{\upvarphi}^{(k)}$. This step is referred as the ``ping" step.
		\item A second step dedicated to the optimization of the transmit waveform, $\boldsymbol{\upvarphi}^{(k+1)}$, given the previously obtained receive waveform $\boldsymbol{\uppsi}^{(k)}$. This step is referred as the ``pong" step.
\end{itemize}
Accordingly, the proposed optimization approach is referred as the POPS algorithm, since it is based on ``ping" and ``pong" steps, to compute, offline, the best transmit and receive waveforms for any transmission impairments.
\begin{algorithm}[H]\caption{Optimization approach}
\begin{algorithmic}\label{secondapproch}
\REQUIRE $\boldsymbol{\upvarphi}^{(0)}$ and introduce a precision parameter ${\epsilon}>0$ and the maximum number of authorized iterations.
\WHILE{ $e^{(\boldsymbol{\upvarphi})}>\epsilon$ or $e^{(\boldsymbol{\uppsi})}>\epsilon$ }
  \STATE Compute the Kernels $\mathrm{\textbf{KS}}_{S(p,{\nu})}^{\boldsymbol{\upvarphi}^{(k)}}$ and $\mathrm{\textbf{KI}}_{S(p,{\nu})}^{\boldsymbol{\upvarphi}^{(k)}}$
  \STATE $\mathrm{\textbf{KIN}}_{S(p,{\nu})}^{\boldsymbol{\upvarphi}^{(k)}}=
  \mathrm{\textbf{KI}}_{S(p,{\nu})}^{\boldsymbol{\upvarphi}^{(k)}}+
  \mathrm{SNR}^{-1}\|\boldsymbol{\upvarphi}^{(k)}\|^2{\textbf I}_{N_{\boldsymbol{\uppsi}}}$
	\STATE $\left[\textbf{U},\boldsymbol{\Lambda}\right]=\mathrm{eig}\left(\mathrm{\textbf{KIN}}_{S(p,{\nu})}^{\boldsymbol{\upvarphi}^{(k)}}\right)$
	\item $\boldsymbol{\Phi}=\boldsymbol{\Lambda}^{-\frac{1}{2}}\textbf{U}^H\mathrm{\textbf{KS}}_{S(p,{\nu})}^{\boldsymbol{\upvarphi}^{(k)}}\textbf{U}\boldsymbol{\Lambda}^{-\frac{1}{2}}$
  \STATE $\left[\textbf{u}_{\mathrm{max}},\lambda_{\mathrm{max}}\right]=\mathrm{eigs}(\boldsymbol{\Phi})$
  \STATE $\boldsymbol{\uppsi}^{(k)}=\frac{\textbf{U}\boldsymbol{\Lambda}^{-\frac{1}{2}}\textbf{u}_{\mathrm{max}}}{\|\textbf{U}\boldsymbol{\Lambda}^{-\frac{1}{2}}\textbf{u}_{\mathrm{max}}\|}$
  \STATE Calculate the Kernels $\mathrm{\textbf{KS}}_{S(-p,-{\nu})}^{\boldsymbol{\uppsi}^{(k)}}$ and $\mathrm{\textbf{KI}}_{S(-p,-{\nu})}^{\boldsymbol{\uppsi}^{(k)}}$
  \STATE $\mathrm{\textbf{KIN}}_{S(-p,-{\nu})}^{\boldsymbol{\uppsi}^{(k)}}=\mathrm{\textbf{KI}}_{S(-p,-{\nu})}^{\boldsymbol{\uppsi}^{(k)}}+\mathrm{SNR}^{-1}\|\boldsymbol{\uppsi}^{(k)}\|^2{\textbf I}_{N_{\boldsymbol{\upvarphi}}}$
	\STATE $\left[\textbf{V},\boldsymbol{\Xi}\right]=\mathrm{eig}\left(\mathrm{\textbf{KIN}}_{S(-p,-{\nu})}^{\boldsymbol{\uppsi}^{(k)}}\right)$
	\STATE $\boldsymbol{\Gamma}=\boldsymbol{\Xi}^{-\frac{1}{2}}\textbf{V}^H\mathrm{\textbf{KS}}_{S(-p,-{\nu})}^{\boldsymbol{\uppsi}^{(k)}}\textbf{V}\boldsymbol{\Xi}^{-\frac{1}{2}}$
  \STATE $\left[\textbf{v}_{\mathrm{max}},\vartheta_{\mathrm{max}}\right]=\mathrm{eigs}(\boldsymbol{\Gamma})$
  \STATE $\boldsymbol{\upvarphi}^{(k+1)}=\frac{\textbf{V}\boldsymbol{\Xi}^{-\frac{1}{2}}\textbf{v}_{\mathrm{max}}}{\|\textbf{V}\boldsymbol{\Xi}^{-\frac{1}{2}}\textbf{v}_{\mathrm{max}}\|}$
\STATE Evaluate  errors $e^{(\boldsymbol{\upvarphi})}=\|\boldsymbol{\upvarphi}^{(k+1)}-\boldsymbol{\upvarphi}^{(k)}\|$ and     $e^{(\boldsymbol{\uppsi})}=\|\boldsymbol{\uppsi}^{(k+1)}-\boldsymbol{\uppsi}^{(k)}\|$
\ENDWHILE
 \end{algorithmic}
\end{algorithm}
\section{Computation of simplified versions of useful and interference kernels}
The complexity of the proposed approach could be categorized with respect to offline and online processing. On the one hand, offline processing, which is the concern of the present paper, is done once and beforehand, with the aim of finding the most suitable codebook size, under a complexity-efficiency compromise perspective, the best representative statistics of the channel and the best corresponding pairs of transmit-receive waveforms. This processing can rely on a standard, general and familiar software, such as MATLAB, and requires a very short time (several minutes) for the optimization of the whole set of codebooks required for all applications or mechanisms (such as CoMP or MBMS, C-RAN and the random access channel). On the other hand, online processing encompasses standard waveform filtering, at both transmit and receive sides, as well as run-of-river channel statistics estimation and transmitter-receiver signaling, to adapt the transmit-receive waveform pair to the changing propagation statistics. Among these tasks, filtering is by far the most computational-resource consuming. Hence, we believe that online processing will mostly be of similar complexity to standard and thoroughly studied FBMC systems.\\
In order to reduce the offline complexity of the Algorithm~\ref{secondapproch}, we next derive simplified expressions of the useful and interference Kernels (equations~(\ref{useful-kernel}) and (\ref{interference-kernel})), taking into account the channel characteristics and the quincunx/hexagonal nature of multicarrier transmission. Despite the fact that the presented paradigm can take into account the cumulative effects of asynchronism, synchronization errors and delay and Doppler spreads, for briefly, we only focus next on natural impairments, brought by the channel. For this, we consider a scattering function with decoupled diffuse Doppler spectral density in the frequency domain, $\alpha(\nu)$, obeying the Jakes model~\cite{Jung}, and a discrete-time multipath-power profile in the time domain, $\beta(p)$, obeying the exponential truncated decaying model. Hence
\begin{equation}\label{modelcanal}S(p,{\nu})=\alpha(\nu)\beta(p),
\end{equation}
with
\begin{equation}\label{6}\alpha(\nu)= \left\{\begin{array}{ll}\frac{1}{\pi
f_D}\frac{1}{\sqrt{1-(\nu/f_D)^2}}\hspace{0.2cm} \mathrm{if} \hspace{0.2cm}\left|\nu\right|<f_D\\
\cr 0 \hspace{2.5cm} \mathrm{if}\hspace{0.2cm} f_D\leq\left|\nu\right|\leq{\frac{1}{2T_s}}
\end{array}\right.
\end{equation}
where $f_D$ is the maximum Doppler frequency shift, and
\begin{equation}\label{6}\beta(p)=\displaystyle{\sum_{k=0}^{K-1}}\pi_k\delta_K(p-p_k),
\end{equation}
with normalized power paths $\pi_k=\frac{1-b}{1-b^{K}}b^k$, $k=0,1,\cdots,K-1$, with $0<b<1$ being the decaying factor and $K$ being the number of contiguous paths in the channel.\\
For the present channel, the useful Kernel can be written as
\begin{equation} \begin{array}{ll}
\mathrm{\textbf{KS}}_{S(p,\nu)}^{\boldsymbol{\upvarphi}}
&=\boldsymbol{\Pi}\odot\displaystyle{\sum_{k=0}^{K-1}}\pi_k
\left(\boldsymbol{\upsigma}_{p_k}(\boldsymbol{\upvarphi})\boldsymbol{\upsigma}_{p_k}(\boldsymbol{\upvarphi})^H\right),
\end{array}
\end{equation}
where $\boldsymbol{\Pi}$ is the Hermitian matrix representing the time autocorrelation function of the channel, whose $(p,q)^{\mathrm{th}}$ entry is given by
$$\left(\boldsymbol{\Pi}\right)_{pq}=\int\alpha(\nu)e^{j2{\pi}{\nu}T_s(p-q)}d\nu=J_0\left(2{\pi}f_DT_s(p-q)\right).$$
As for the second interference Kernel, for hexagonal lattices, we can write
\begin{equation}\label{kernelinfini} \begin{array}{lllllllll}
\mathrm{\textbf{KI}}_{S(p,\nu)}^{\boldsymbol{\upvarphi}}=\hspace{0.01cm}{_{\infty}\hspace{-0.04cm}{\mathrm{\textbf{K}}}_{S(p,{\nu})}^{\boldsymbol{\upvarphi}}-\mathrm{\textbf{KS}}_{S(p,\nu)}^{\boldsymbol{\upvarphi}}},
\end{array}
\end{equation}
where $_{\infty}\hspace{-0.04cm}\mathrm{\textbf{K}}_{S(p,{\nu})}^{\boldsymbol{\upvarphi}}$ is a new ``infinite" Kernel, defined by
\begin{equation}
_{\infty}\hspace{-0.04cm}\mathrm{\textbf{K}}_{S(p,{\nu})}^{\boldsymbol{\upvarphi}}=\hspace{0.08cm}_{\infty}\mathrm{\textbf{KE}}_{S(p,{\nu})}^{\boldsymbol{\upvarphi}}+\hspace{0.08cm}_{\infty}\mathrm{\textbf{KO}}_{S(p,{\nu})}^{\boldsymbol{\upvarphi}},
\end{equation}
where
\begin{equation} \hspace{0.08cm}_{\infty}\mathrm{\textbf{KE}}_{S(p,{\nu})}^{\boldsymbol{\upvarphi}}=\boldsymbol{\Omega}_{\mathrm{e}}\odot\displaystyle{\sum_{k=0}^{K-1}}\pi_k\displaystyle{\sum_{n}}\left(
\boldsymbol{\upsigma}_{p_k+nN}(\boldsymbol{\upvarphi}){\odot}\left(\boldsymbol{\upsigma}_{p_k+nN}(\boldsymbol{\upvarphi})\right)^H\right)
\end{equation}
is an ``even" infinite Kernel, with
\begin{equation}
\boldsymbol{\Omega}_{\mathrm{e}}=\boldsymbol{\Pi}\odot\left(\displaystyle{\sum_{\delta=0}^{Q/2-1}}e^{j4{\pi}\delta(p-q)/Q}\right)_{(p,q)\in\mathbb{Z}^2}
\end{equation}
being a Hermitian matrix, whose $(p,q)^{\mathrm{th}}$ entry is given by
\begin{equation}
\left(\boldsymbol{\Omega}_{\mathrm{e}}\right)_{pq}=\left\{\begin{array}{ll}
\frac{Q}{2}J_0(2{\pi}f_DT_s(p-q)) \hspace{0.2cm} \mathrm{if} \hspace{0.2cm} (p-q)\hspace{0.1cm}\mathrm{mod}\hspace{0.1cm}Q/2=0 \\
\cr 0 \hspace{0.2cm} \mathrm{otherwise},
\end{array}\right.
\end{equation}
and
\begin{equation}
\hspace{0.08cm}_{\infty}\mathrm{\textbf{KO}}_{S(p,{\nu})}^{\boldsymbol{\upvarphi}}=\boldsymbol{\Omega}_{\mathrm{o}}\odot\displaystyle{\sum_{k=0}^{K-1}}\pi_k\displaystyle{\sum_{n}}\left(\boldsymbol{\upsigma}_{p_k+(n+\frac{1}{2})N}(\boldsymbol{\upvarphi}){\odot}\left(\boldsymbol{\upsigma}_{p_k+(n+\frac{1}{2})N}(\boldsymbol{\upvarphi})\right)^H\right)
\end{equation}
is an ``odd" infinite Kernel, with
\begin{equation}
\boldsymbol{\Omega}_{\mathrm{o}}=\boldsymbol{\Pi}\odot\left(\displaystyle{\sum_{\delta=0}^{Q/2-1}}e^{j2{\pi}(2\delta+1)(p-q)/Q}\right)_{(p,q)\in\mathbb{Z}^2}
\end{equation}
being a Hermitian matrix whose $(p,q)^{\mathrm{th}}$ entry is given by
\begin{equation}
\left(\boldsymbol{\Omega}_{\mathrm{o}}\right)_{pq}=\left\{\begin{array}{ll}
-\frac{Q}{2}J_0(2{\pi}f_DT_s(p-q)) \hspace{0.2cm} \mathrm{if} \hspace{0.2cm} (p-q)\hspace{0.1cm}\mathrm{mod}\hspace{0.1cm}Q=Q/2 \\
\cr \frac{Q}{2}J_0(2{\pi}f_DT_s(p-q)) \hspace{0.2cm} \mathrm{if} \hspace{0.2cm} (p-q)\hspace{0.1cm}\mathrm{mod}\hspace{0.1cm}Q=0 \\
\cr 0 \hspace{0.2cm} \mathrm{otherwise}.
\end{array}\right.
\end{equation}

\begin{figure}[H]
\centering
    \subfigure[ Useful Kernel]
   {\hspace{1cm}\includegraphics[width=0.42\textwidth]{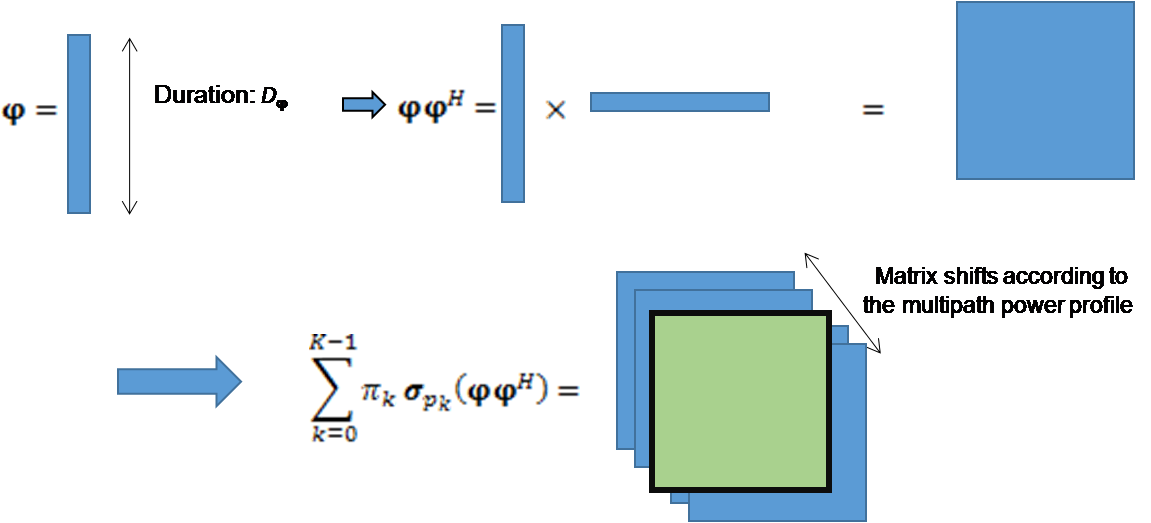}\label{calculkernelpopsa}
}\vfill\vspace{-0.3cm}

    \subfigure[``Even" infinite Kernel]
  {\hspace{-1.6cm}\includegraphics[width=0.65\textwidth]{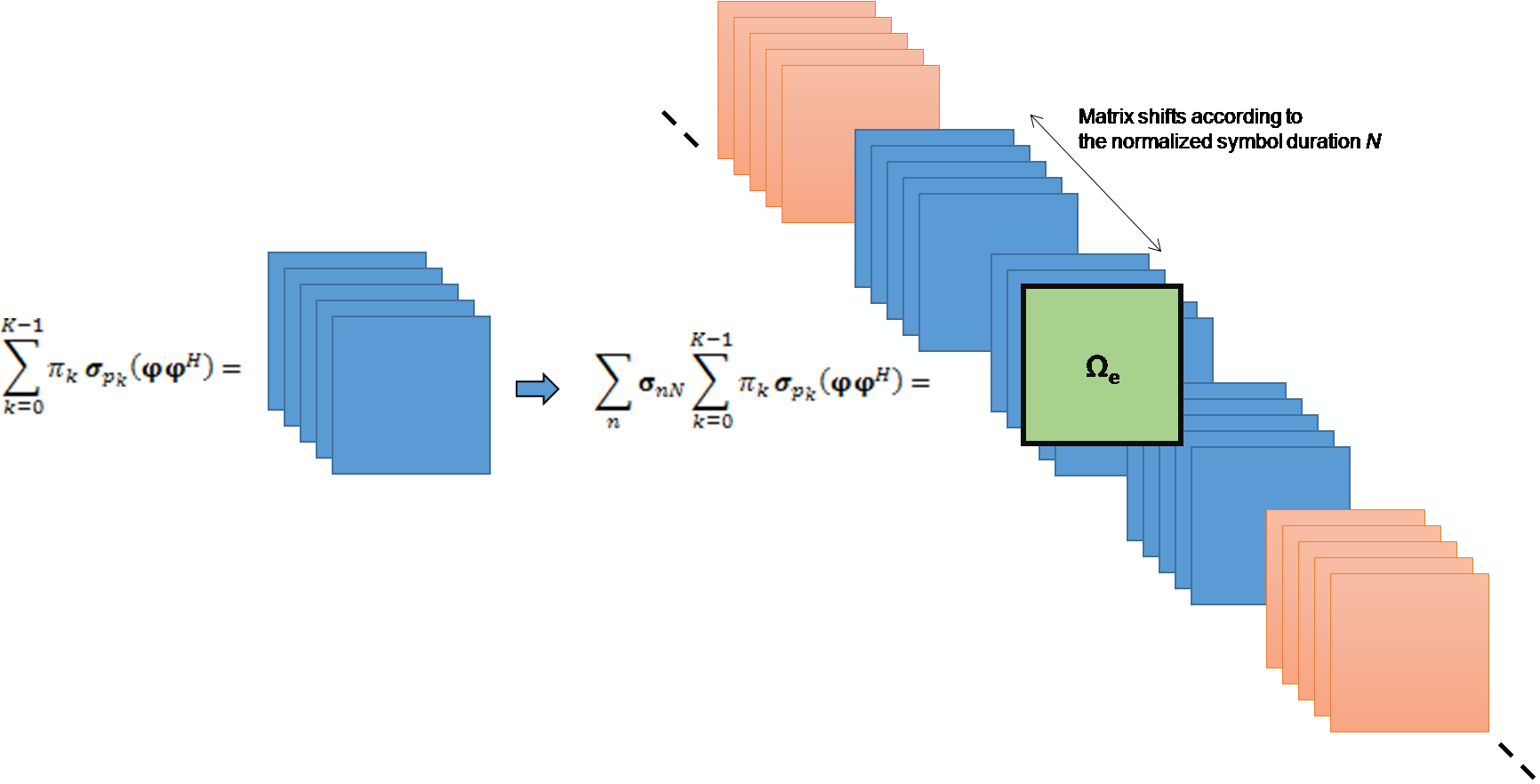}\label{calculkernelpopsb}
   }\vfill\vspace{-0.3cm}

      \subfigure[``Odd" infinite Kernel]
    {
  \hspace{-2.4cm}\includegraphics[width=0.63\textwidth]{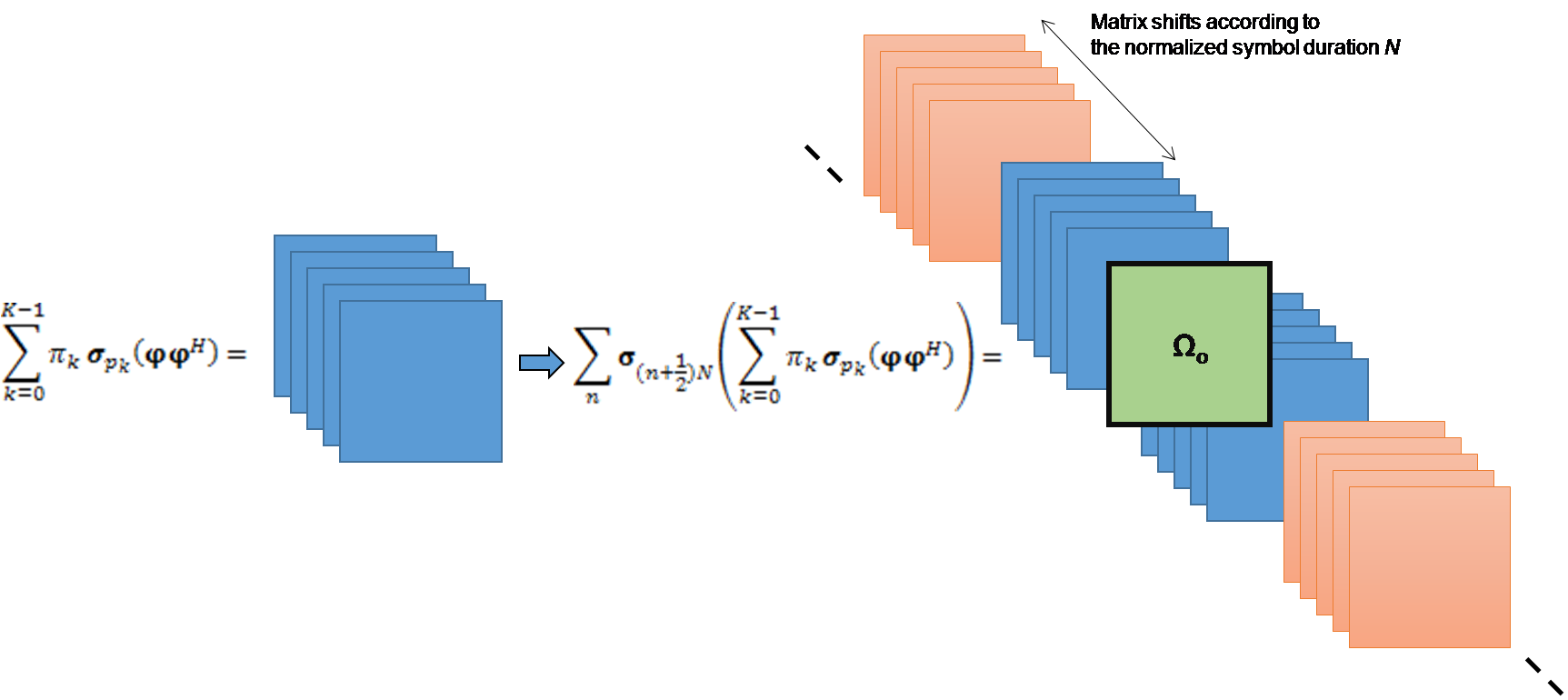}\label{calculkernelpopsc}

    }
 \caption{Illustration of the approach for calculating the useful and interference Kernels. (The pink squares of ``even" and ``odd" infinite Kernels do not overlap with the selection window and therefore do not contribute to the determination of the infinite Kernel.)}
\label{fig:calculkernelpops}
\end{figure}

Notice that the ``even" infinite Kernel accounts for the contributions of even subcarriers presenting a zero time shift with respect to the $0^{\mathrm{th}}$ subcarrier and the ``odd" infinite Kernel accounts for the contributions of the odd subcarriers presenting a half-symbol-period time shift with respect to the $0^{\mathrm{th}}$ subcarrier. Also notice that the matrices, $\boldsymbol{\Pi}$, $\boldsymbol{\Omega}_{\mathrm{e}}$ and $\boldsymbol{\Omega}_{\mathrm{o}}$, are calculated once for the whole optimization process and are kept unchanged from one iteration to another.\\
As illustrated in Fig.~\ref{fig:calculkernelpops}, we next propose an easy and efficient method to simplify the numerical calculation of the previous Kernels, $\mathrm{\textbf{KS}}_{S(p,\nu)}^{\boldsymbol{\upvarphi}}$ and $\mathrm{\textbf{KI}}_{S(p,\nu)}^{\boldsymbol{\upvarphi}}$. The calculation of these Kernels is performed at the ``ping" step of each iteration $k$ for the optimization of the receive waveform $\boldsymbol{\uppsi}$. Likewise, following the same process as in Fig.~\ref{fig:calculkernelpops}, the ``pong" step of each iteration banks on a determination of both Kernels, $\mathrm{\textbf{KS}}_{S(-p,-\nu)}^{\boldsymbol{\uppsi}}$ and $\mathrm{\textbf{KI}}_{S(-p,-\nu)}^{\boldsymbol{\uppsi}}$, by replacing the transmit waveform $\boldsymbol{\upvarphi}$ by the receive waveform, $\boldsymbol{\uppsi}$, and considering instead the inverse in time and frequency of channel scattering function. According to  Fig.~\ref{fig:calculkernelpops}\subref{calculkernelpopsa}, we start by multiplying the transmit waveform $\boldsymbol{\upvarphi}$ by its Hermitian transpose $\boldsymbol{\upvarphi}^H$. The resulting matrix is shifted according to the multipath power profile of the channel and then entrywise multiplied with matrix $\boldsymbol{\Pi}$. Subsequently, the useful Kernel $\mathrm{\textbf{KS}}_{S(p,\nu)}^{\boldsymbol{\upvarphi}}$ is obtained using a selection window, in gap with that of the transmit waveform, to take into account the causality of the channel. As shown in Fig.~\ref{fig:calculkernelpops}\subref{calculkernelpopsa}, Fig.~\ref{fig:calculkernelpops}\subref{calculkernelpopsb} and Fig.~\ref{fig:calculkernelpops}\subref{calculkernelpopsc}, this selection window, which is common to all involved Kernels, plays a key role in the determination of the SINR as a ratio of two quadratic forms in  $\boldsymbol{\uppsi}$. Because of the finite durations of  $\boldsymbol{\upvarphi}$  and  $\boldsymbol{\uppsi}$, the relative position of the the time window of $\boldsymbol{\uppsi}$, with respect to the time window of $\boldsymbol{\upvarphi}$, becomes crucial and decisive in the determination of the achievable SINR. Typically, it is optimized by moving the $\boldsymbol{\uppsi}$ window around the average delay incurred by the channel, in steps of the sampling period $T_s$, and finding the best achievable SINR. \\
Now, in order to compute the elementary contributions to the Kernel $\hspace{0.08cm}_{\infty}\mathrm{\textbf{K}}_{S(p,{\nu})}^{\boldsymbol{\upvarphi}}$, we must compute the ``even" and ``odd" infinite Kernels. As shown in Fig.~\ref{fig:calculkernelpops}\subref{calculkernelpopsb}, the ``even" infinite Kernel, $\hspace{0.08cm}_{\infty}\mathrm{\textbf{KE}}_{S(p,{\nu})}^{\boldsymbol{\upvarphi}}$ is obtained by regularly and periodically shifting the shifted matrix $\sum_{k=0}^{K-1}\pi_k
\left(\boldsymbol{\upsigma}_{p_k}(\boldsymbol{\upvarphi})\right.$ $\left.\boldsymbol{\upsigma}_{p_k}(\boldsymbol{\upvarphi})^H\right)$ by increments of the normalized symbol duration $N$ and multiplying the final result with matrix $\boldsymbol{\Omega}_{\mathrm{e}}$. Also, as shown in Fig.~\ref{fig:calculkernelpops}\subref{calculkernelpopsc}, the ``odd" interference Kernel, $\hspace{0.08cm}_{\infty}\mathrm{\textbf{KO}}_{S(p,{\nu})}^{\boldsymbol{\upvarphi}}$, is obtained by regularly and periodically shifting the shifted matrix $\sum_{k=0}^{K-1}\pi_k
\left(\boldsymbol{\upsigma}_{p_k}(\boldsymbol{\upvarphi})\boldsymbol{\upsigma}_{p_k}(\boldsymbol{\upvarphi})^H\right)$ by integer multiples of $N$ samples starting from an initial shift by $N/2$ samples in time and, finally, multiplication this result with matrix $\boldsymbol{\Omega}_{\mathrm{o}}$. The infinite Kernel is obtained by selecting the window from the optimum position after summing both ``even" and ``odd" infinite Kernels. \\
In order to determine the interference Kernel $\mathrm{\textbf{KI}}_{S(p,\nu)}^{\boldsymbol{\upvarphi}}$, we subtract the useful Kernel $\mathrm{\textbf{KS}}_{S(p,\nu)}^{\boldsymbol{\upvarphi}}$ from the infinite Kernel  $_{\infty}\mathrm{\textbf{K}}_{S(p,{\nu})}^{\boldsymbol{\upvarphi}}$. Since the transmit and receive waveforms have finite durations, only a finite number of the shifts of  $\sum_{k=0}^{K-1}\pi_k
\left(\boldsymbol{\upsigma}_{p_k}(\boldsymbol{\upvarphi})\boldsymbol{\upsigma}_{p_k}(\boldsymbol{\upvarphi})^H\right)$ actually contribute to the full determination of the selection window needed in the determination of the SINR.
\section{NUMERICAL RESULTS}
In this section, we numerically evaluate the performance of FBMC with waveforms optimized for hexagonal lattices and operating over highly time and frequency dispersive channels, characterized by
channel scattering functions obeying~\eqref{modelcanal}. In a first bunch of numerical results, we fix the number of subcarriers to $Q=128$ and the spread factor to $B_dT_m=10^{-2}$. In a second bunch of numerical results, we vary these values to study the performance of POPS-FBMC as a function of $Q$ and $B_dT_m$. According to past experimental characterizations of the small-scale part of the propagation channel, the maximum reported delay spread, $T_m$, never exceeds a few microseconds~\cite{Greenstein}, while the maximum Doppler frequency, $B_d$, never goes above a few hundred Hertz~\cite{gpp} for a high mobility. As a consequence, the channel spread factor of $10^{-2}$, is typically one to two order of magnitude larger that practically reported values. Camping on these large values of the channel spread not helps us tackle a worst case situation but also take into account other synchronization and asynchronism imperfections. In order to have the best value of the SINR, we always go through a preliminary determination of the best balance between $F$ and $T$ with respect to $B_d$ and $T_m$ respectively, while observing a given lattice density constraint $\Delta = 1/FT$. In this determination, we don't care about the optimum initialization of the POPS algorithm and stick to the Gaussian function, as an initialization of the transmit waveform, which is also the first Hermite function, that offers the best localization on the time-frequency plane. Moreover, for simplicity sake and without loss of generality, we choose $N_{\boldsymbol{\upvarphi}}$ and $N_{\boldsymbol{\uppsi}}$ as integer multiples of $N$ and assume a common duration, $D=D_{\boldsymbol{\upvarphi}}=D_{\boldsymbol{\uppsi}}$, for the transmitter and receiver waveforms, although these conditions are not necessary and the general case could be treated as well. We compare the obtained numerical results to two benchmarks, with the first being conventional OFDM and the second being POPS-FBMC with rectangular lattices, with the common time and frequency shifts being respectively $t_{mn}=nT$ and $f_{mn}=mF$.
\subsection{SINR and SIR performances of POPS-FBMC on quincunx/hexagonal lattices}
\begin{figure}[H]
       \centering
\includegraphics[width=0.7\textwidth]{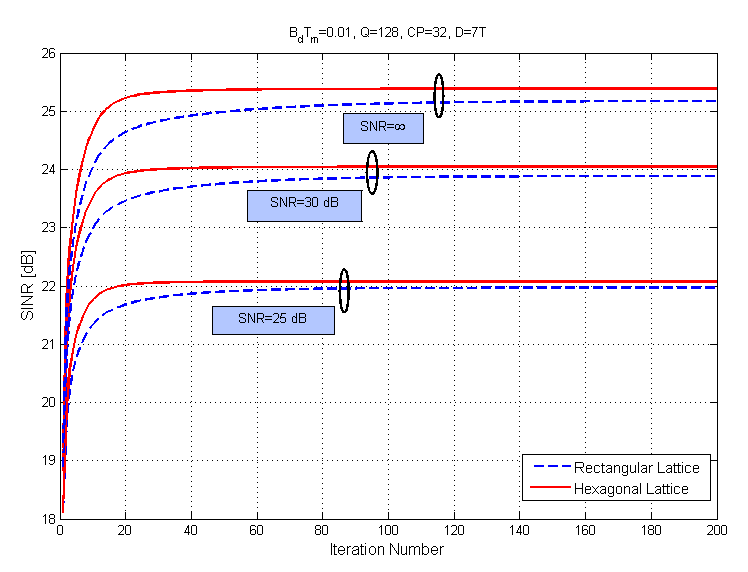}
    \caption{Evolution of the achievable SINR as a function of the POPS algorithm iterations, for $B_dT_m=0.01$, $Q=128$, $\textit{\mbox{CP}}=32$, $D=7T$, $FT=1.25$ and $\mbox{SNR}=25\hspace{0.1cm}\mbox{dB}$, $30\hspace{0.1cm}\mbox{dB}$ and infinity.}
    \label{SINR-iteration}
\end{figure}

\begin{figure}[H]
    \centering
\includegraphics[width=0.7\textwidth]{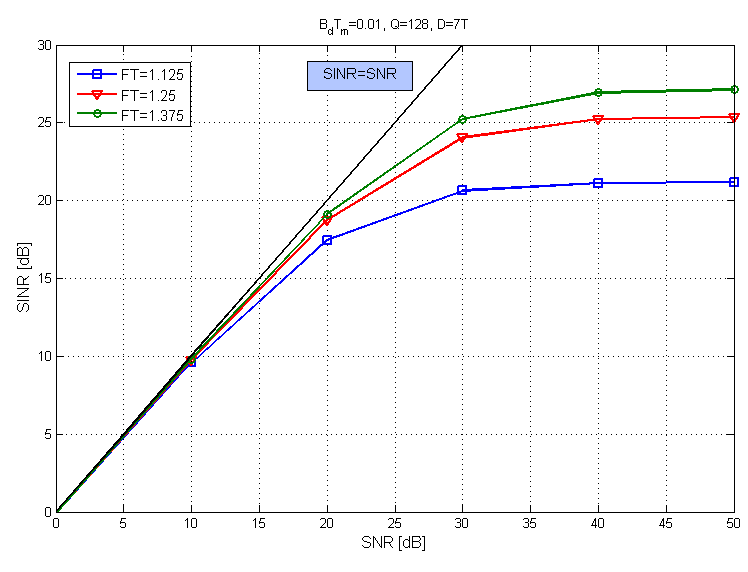}
    \caption{Optimized SINR versus different values of SNR, for $B_dT_m=0.01$, $Q=128$ and $D=7T$.}
        \label{SINR-snr}
\end{figure}

\begin{figure}[H]
    \centering
\includegraphics[width=0.85\textwidth]{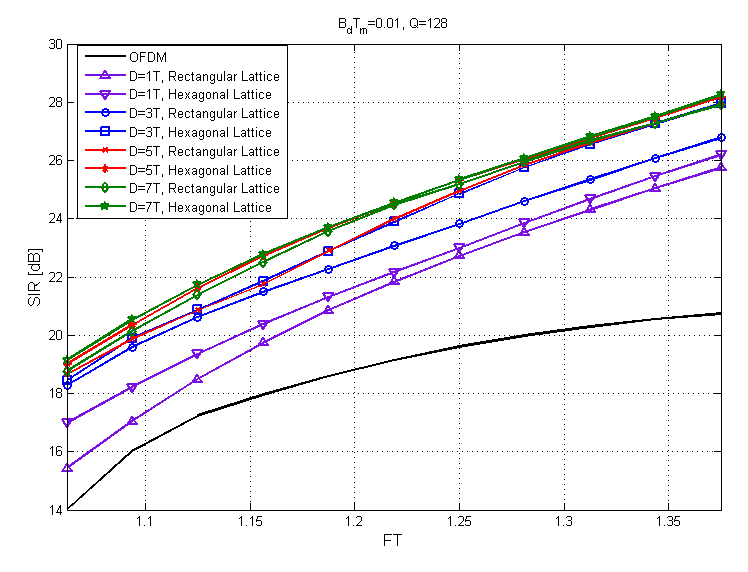}
   \caption{Optimized SIR versus $FT$, for $B_dT_m=0.01$ and $Q=128$.}
\label{SIR-FT}
\end{figure}

\begin{figure}[H]
    \centering
\includegraphics[width=0.85\textwidth]{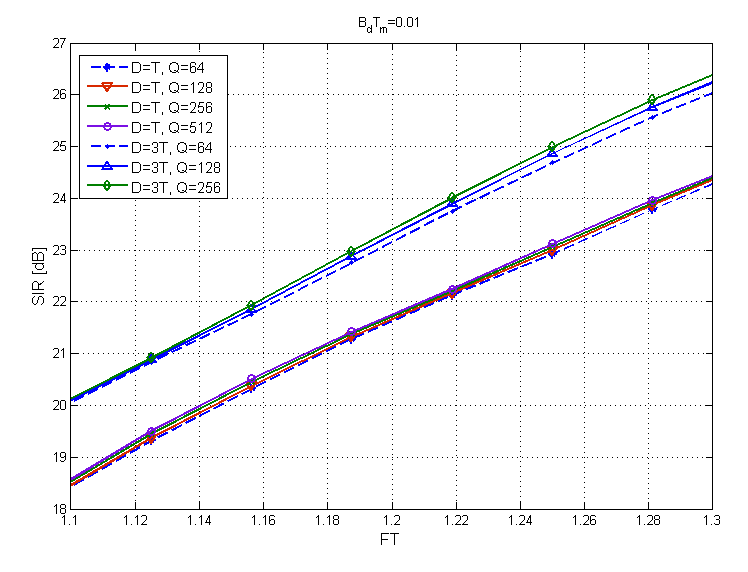}
    \caption{Optimized SIR versus $FT$, for different number of subcarriers $Q$, $B_dT_m=0.01$ and $D=T$ and $3T$.}
\label{SIR-samples}
\end{figure}

\begin{figure}[H]
    \centering
\includegraphics[width=0.85\textwidth]{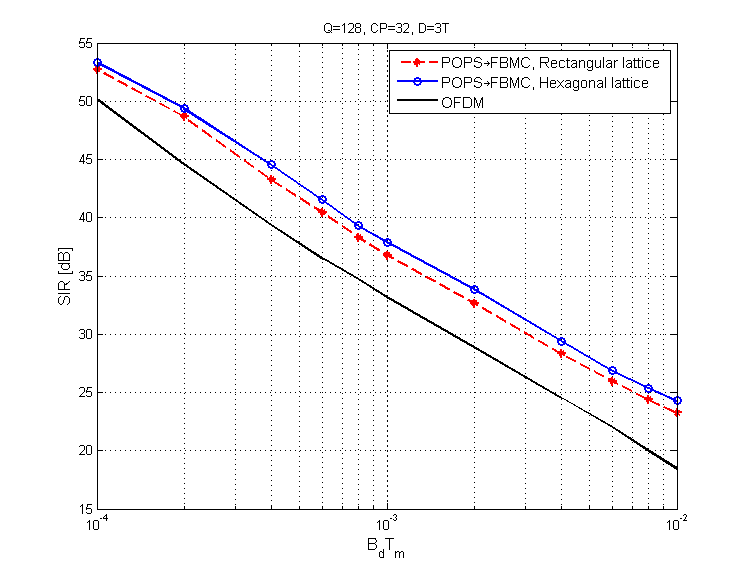}
    \caption{Optimized SIR as a function of the channel spread factor $B_dT_m$, for $Q=128$, $FT=1.25$ and $D=3T$.}
\label{SIR-bdtm}
\end{figure}

In Fig.~\ref{SINR-iteration}, we present the evolution of the SINR as a function of the number of iterations, for hexagonal and rectangular lattices, over a time-frequency dispersive AWGN channel, when $\mbox{SNR}=25\hspace{0.1cm}\mbox{dB}$, $30\hspace{0.1cm}\mbox{dB}$ and infinity, for a lattice density $\Delta=\frac{1}{FT}=0.8$ ($FT=1.25$) and a waveform duration $D=7T$. This figure shows that FBMC, with quincunx/hexagonal lattices, offers a slight improvement in SINR
with respect to FBMC with rectangular lattices, for the different considered values of the $\mbox{SNR}$. We also notice the convergence of the SINR to the SNR over the a time-frequency dispersive AWGN channel, for both rectangular and hexagonal lattices. To better illustrate this result, we draw, in Fig.~\ref{SINR-snr}, the behavior of the optimized SINR as a function of the SNR for a hexagonal multicarrier system, for different values of the lattice density ($\Delta=0.7$, $0.8$ and $0.9$) and a waveform duration $D=7T$. This figure shows that, at low SNR, the optimized SINR is very close to the SNR value. However, at high SNR, the optimized SINR converges to the optimized SIR value obtained for a noiseless channel (i.e. $N_0=0$). Fig.~\ref{SIR-FT} shows the SIR evolution as a function of
$FT$ for optimized systems, on hexagonal and rectangular lattices, operating over time-frequency dispersive
noiseless channels. A comparison is also made with the conventional OFDM system. Different values of the waveform duration ($D=T$, $3T$, $5T$ and $7T$) are considered. Fig.~\ref{SIR-FT} shows an SIR enhancement for the rectangular and hexagonal multicarrier transmission systems when the waveform duration increases or the lattice density decreases. It also shows that the SIR obtained with $D=7T$ slightly outperforms the one obtained with $D=5T$. Fig.~\ref{SIR-FT} also shows that the POPS-FBMC on hexagonal lattices offers a strong improvement in SIR with respect to the conventional OFDM and a maximum improvement of $1\hspace{0.1cm}\mbox{dB}$ with respect to the optimized system on rectangular lattices, depending on waveform duration and lattice density. For example, when $\Delta=0.8$ and $D=3T$, a $5\hspace{0.1cm}\mbox{dB}$ (respectively, $4\hspace{0.1cm}\mbox{dB}$)  gain is obtained with the optimized system on hexagonal (respectively, rectangular) lattices, relatively to conventional OFDM systems. Fig.~\ref{SIR-samples} illustrates the impact of an increase of the number of subcarriers, $Q$, on the SIR behavior of the optimized POPS-FBMC with hexagonal lattices. Two values of the waveform duration, $D=T$ and $D=3T$, are considered. We observe a slight improvement in SIR when $Q$ increases.
For a further characterization of the POPS-FBMC on hexagonal lattices, we show in Fig.~\ref{SIR-bdtm}
the behavior of the SIR versus the channel spread factor $B_dT_m$, for a lattice density $\Delta=0.8$ and a waveform duration $D=3T$. We also add in this figure the SIR obtained with a POPS-FBMC on rectangular lattices and a conventional OFDM. We observe a significant enhancement in SIR when the spread factor decreases for the different systems. We also note that POPS-FBMC on hexagonal lattices always outperforms POPS-OFDM on rectangular lattices and conventional OFDM, for all values of $B_dT_m$. More importantly, we observe that the SIR performance increases with hexagonal-lattice FBMC, with respect to the rectangular-lattice FBMC, when the channel dispersion $B_dT_m$ increases. For example, when $B_dT_m=10^{-2}$, a $1\hspace{0.1cm}\mbox{dB}$ gain is obtained with the optimized system on hexagonal lattices, relatively to the optimized system with rectangular lattices, while a $0.5\hspace{0.1cm}\mbox{dB}$ gain is obtained when $B_dT_m=10^{-4}$.
\subsection{Optimized waveforms characterization on hexagonal lattices}
\begin{figure}[H]
\centering
  \includegraphics[width=0.85\textwidth]{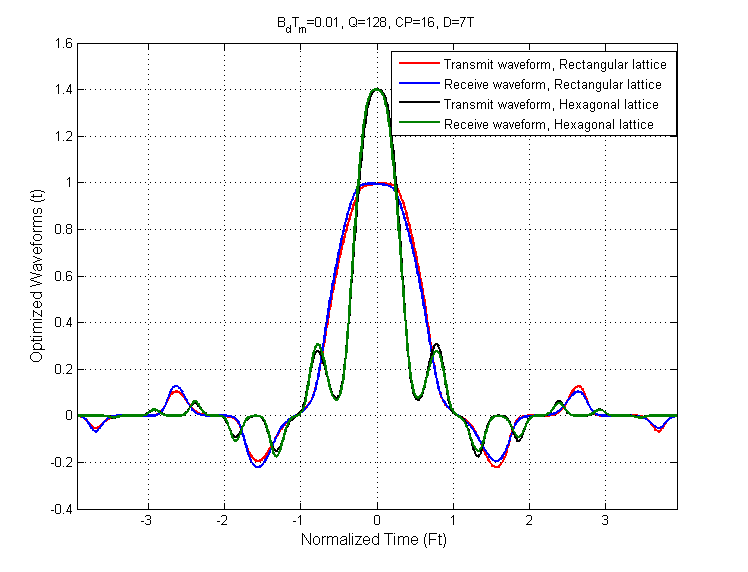}
\caption{Optimized transmit and receive waveforms on the hexagonal and rectangular lattices, for $B_dT_m=0.01$, $Q=128$, $D=7T$ and $FT=1.125$.}
\label{waveformlattice}
\end{figure}

\begin{figure}[H]
\centering
    \subfigure[]
  {\includegraphics[width=0.49\textwidth]{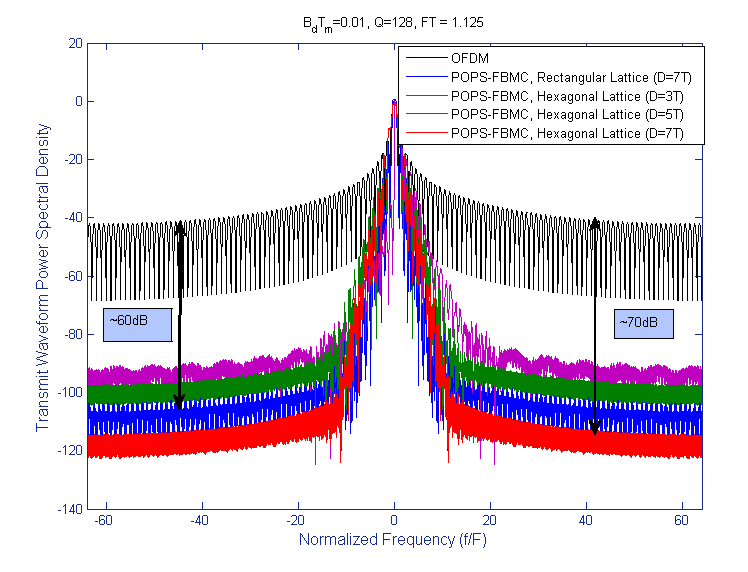}\label{spectre1}
}\hfill
    \subfigure[]
  {\includegraphics[width=0.49\textwidth]{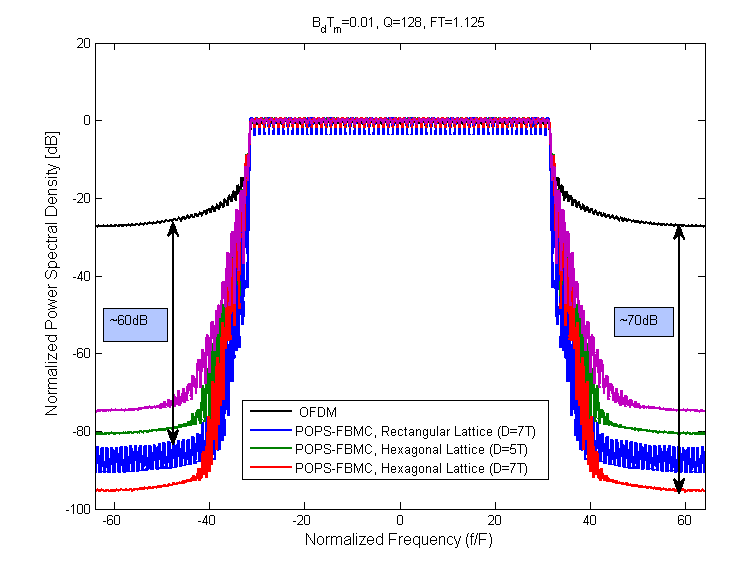}\label{spectre2}
   }
      \caption{Normalized power spectral density, for $B_dT_m=0.01$, $Q=128$ and $FT=1.125$. (a) One subcarrier. (b) Aggregation of $65$ contiguous subcarriers.}
\label{fig:fctspectre}
\end{figure}

In Fig.~\ref{waveformlattice}, we present the temporal evolution of the optimum transmit and receive waveforms
corresponding to the optimized SIR, when $\mbox{SNR}=\infty$, for $\Delta=0.9$, $D=7T$ and both hexagonal and rectangular lattices. For both lattices, we observe that the optimized receive waveform matches perfectly the temporal inverse of the optimized transmit waveform. Using alternative expressions of the SINR as generalized Rayleigh quotients in the receive waveform $\boldsymbol{\uppsi}$, we can prove that this characteristic is theoretically valid when the optimum pair of transmit/receive waveforms is unique. We also note that the main lobes of the optimized transmit and receive waveforms obtained on hexagonal lattices are more concentrated around the origin, compared to the ones on rectangular lattices. This better localization justifies the SIR
performance improvement on hexagonal lattices. \\
In Fig.~\ref{fig:fctspectre}, we depict the power spectral densities (PSD) of the optimized transmit waveform  for the hexagonal lattice (Fig.~\ref{spectre1}) as well as for $65$ contiguous subcarriers (Fig.~\ref{spectre2}), for different waveform durations and $\Delta=0.9$. A comparison is made with the PSD, of the conventional OFDM waveform and the POPS-optimized waveform for rectangular lattices, for $D=7T$. On the one hand, we note that the OOB power leakage of conventional OFDM is very important, requiring an insertion of large guard bands between the subcarriers of different users, mainly at the uplink when the transmission is asynchronous and the receive power is generally unbalanced between the different communications. On the second hand, we note that the optimized waveform on hexagonal lattices reduces the OOB emission with respect to the conventional OFDM, especially when $D$ increases. For example, a gain of $70 \hspace{0.1cm}\mbox{dB}$ is brought by the POPS-FBMC on hexagonal lattices with $D=7T$, compared to the conventional OFDM system. On the third hand, we point out a $10\hspace{0.1cm}\mbox{dB}$ reduction in OOB emission when we consider the hexagonal layout instead of the rectangular one. This reduction in OOB emission minimizes the interference into neighbouring bands and reduces the probability of viloating spectral masks imposed by some standards.
\subsection{Robustness characterization of POPS-FBMC on quincunx/hexagonal lattices}
\begin{figure}[H]
\centering
  \subfigure[]
 { \includegraphics[width=0.485\textwidth]{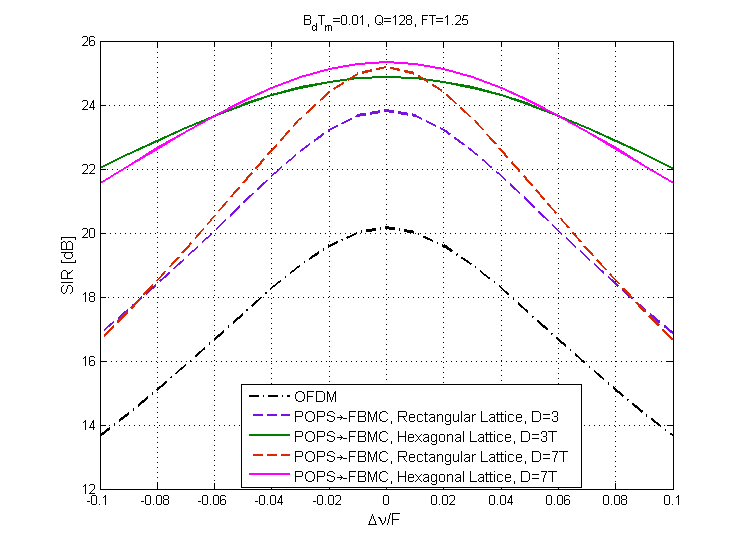}\label{sync-freq}}
\hfill
   \subfigure[]
        { \includegraphics[width=0.485\textwidth]{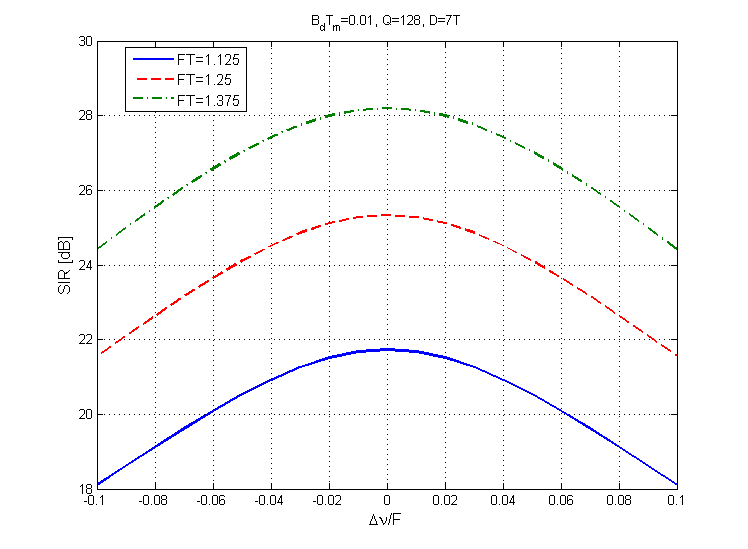}\label{sync-freq-ft}}
 \caption{Sensitivity to synchronization errors in frequency, for $B_dT_m=0.01$, $Q=128$, (a) hexagonal and rectangular lattices, $FT=1.25$, $D=3T$ and $7T$. (b) hexagonal lattices and three values of $FT$.}
\end{figure}

\begin{figure}[H]
\centering
  \subfigure[]
 { \includegraphics[width=0.48\textwidth]{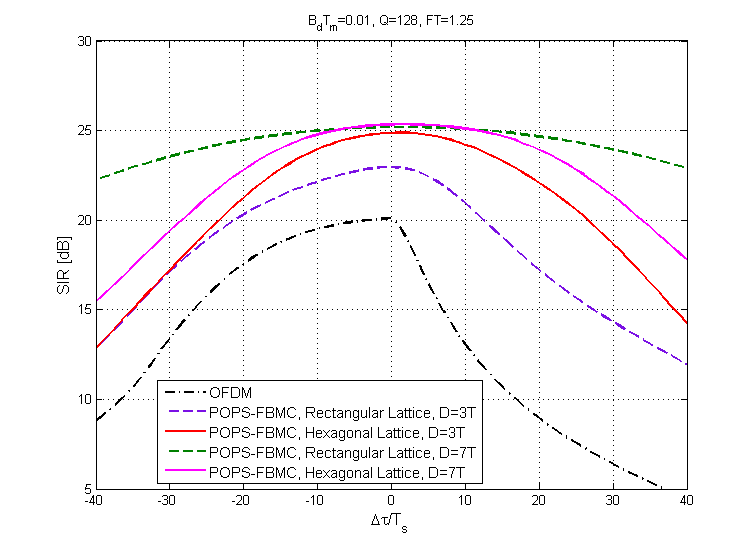}\label{sync-temps}}
\hfill
   \subfigure[]
        { \includegraphics[width=0.48\textwidth]{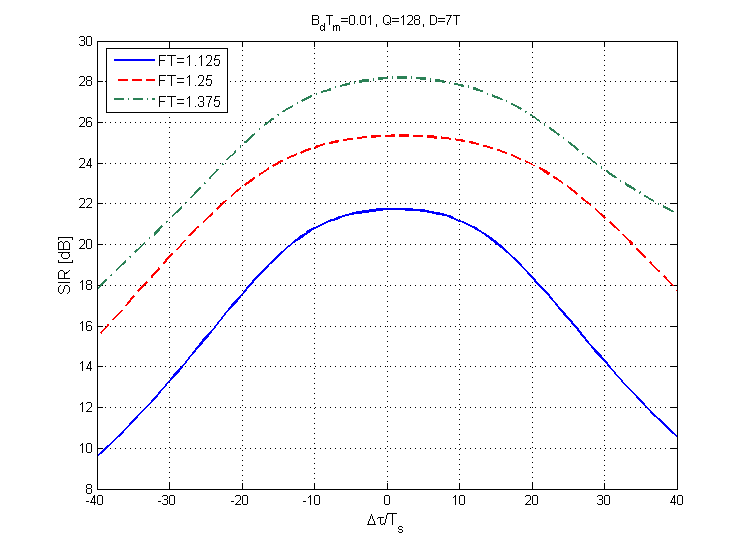}\label{sync-temps-ft}}
 \caption{Sensitivity to synchronization errors in time, for $B_dT_m=0.01$, $Q=128$, (a) hexagonal and rectangular lattices, $FT=1.25$, $D=3T$ and $7T$. (b) hexagonal lattices and three values of $FT$.}
\end{figure}

\begin{figure}[H]
    \centering
\includegraphics[width=1\textwidth]{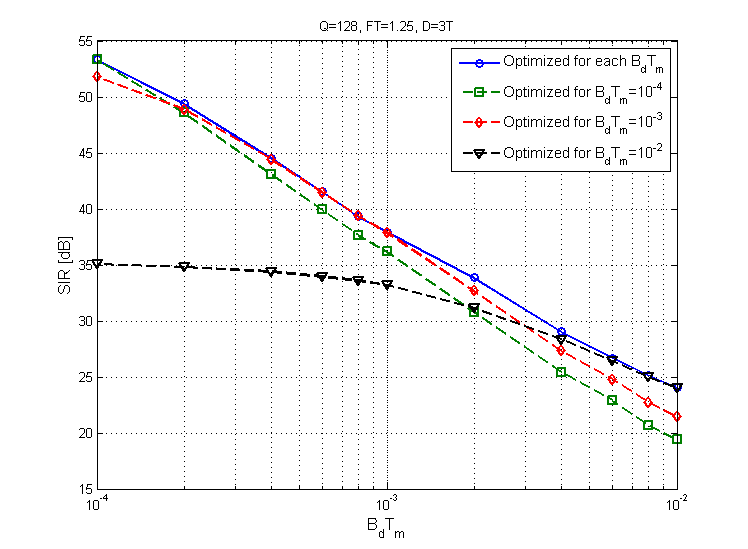}
    \caption{Sensitivity to an estimation error on $B_dT_m$, for $Q=128$, $FT=1.25$ and $D=3T$.}
     \label{sync-err-bdtm}
\end{figure}

In this subsection, we characterize the robustness of POPS-FBMC with respect to its sensitivity to frequency and time synchronization and channel spread factor estimation errors. In Fig.~\ref{sync-freq}, we compare the sensitivity to frequency synchronization errors of optimized systems on both hexagonal and rectangular lattices to that of conventional OFDM. The channel is assumed to be noiseless, meaning that $\mbox{SNR}=\infty$. In this figure, we consider a normalized frequency synchronization error $\Delta\nu/F$ varying between $-0.1$ and $0.1$, a lattice density $\Delta=0.8$ and waveform durations $D=3T$ and $7T$. As it is expected, the SIR degrades with the increase of the frequency synchronization error. More importantly, we notice that POPS-FBMC on hexagonal lattices is less sensitive to frequency synchronization errors than POPS-FBMC on rectangular lattices. For example, a gain in terms of SIR of about $6\hspace{0.1cm}\mbox{dB}$ can be reached by using hexagonal-lattice waveforms instead of rectangular-lattice waveforms, for a waveform duration $D=7T$ and $\Delta\nu/F={\pm}0.1$. Furthermore, Fig.~\ref{sync-freq-ft} shows the sensitivity of the optimized POPS-FBMC on hexagonal lattices to frequency synchronization errors, for different values of the lattice density ($\Delta=0.7$, $0.8$ and $0.9$) and a waveform duration $D=7T$. Notice that the hexagonal-lattice FBMC is less sensitive to frequency synchronization
errors than rectangular-lattice FBMC when $FT$ increases.\\
The sensitivity of POPS-FBMC on hexagonal and rectangular lattices to time synchronization errors is shown in Fig.~\ref{sync-temps}, when the normalized time synchronization error $\Delta\tau/T_s$ varies between $-40$ and $40$, for a lattice density $\Delta=0.8$, waveform durations $D=3T$ and $7T$, and a noiseless channel. Contrarily to the sensitivity to frequency synchronization errors, the POPS-FBMC with hexagonal lattices is more sensitive to ISI over a time dispersive channel than the POPS-FBMC with rectangular lattices, especially for large waveform durations. These extra sensitivity results are explained both by the narrow gap between the successive time shifts and the higher optimal value of $T_m/T$ for hexagonal lattices as compared to rectangular ones. Nevertheless, both hexagonal and rectangular optimized systems outperform the conventional OFDM even when the time synchronization error is very important. Moreover, we show in Fig.~\ref{sync-temps-ft} the sensitivity to time synchronization errors of the hexagonal-lattice POPS-FBMC, for different values of $FT$ when $D=7T$. It is clear that POPS-FBMC with hexagonal lattices becomes less sensitive to time synchronization errors when $FT$ increases. \\
Fig.~\ref{sync-err-bdtm} illustrates the sensitivity of POPS-FBMC, with hexagonal lattices, to spread factor estimation
errors, with actual values of $B_dT_m$ ranging between $10^{-4}$ and $10^{-2}$, for a lattice density $\Delta=0.8$ and a common waveform duration $D=3T$. We show the behavior of the achievable SIR for each of the optimal waveforms, obtained for $\left(B_dT_m\right)_1=10^{-4}$, $\left(B_dT_m\right)_2=10^{-3}$ and $\left(B_dT_m\right)_3=10^{-2}$, as a function of $B_dT_m$. In this figure, we prove the robustness of our optimized waveforms to estimation errors in $B_dT_m$. We observe that the SIR performance degradation, compared to the optimum case, becomes large when the actual $B_dT_m$ becomes smaller than the $B_dT_m$ used for optimization of the transmit-receive waveform pair, than when the actual $B_dT_m$ becomes larger. Fig.~\ref{sync-err-bdtm} is also a good illustration of how the AWC concept can be introduced in 5G systems and beyond. The idea is to determine the codebook, of required waveform pairs, suitable for an efficient adaptation of the transmit and receive waveforms to the channel statistics, while having an acceptable degradation with regards to the case of a perfect knowledge of $B_dT_m$. More precisely, a finite number of waveform pairs could be optimized offline for several well-chosen values of $B_dT_m$. Then, each pair is used online, during effective transmission, for a range of values of $B_dT_m$ around the underlying value of its optimization to outperforms other pairs. As a consequence, the global obtained performance corresponds to the maximum curves of the performances of all pairs in the codebook. Compared to the ideal, yet unrealistic case, where the best pair for the current value of $B_dT_m$ is used, the performance degradation brought by the use of the codebook is conditioned by the gap between the maximum curve and the ideal curve of the SIR and the \textit{a priori} statistics of the taken values of $B_dT_m$. Therefore, the targeted values of $B_dT_m$, underlying the determination of the codebook, must be well chosen to minimize the average loss in performance for a given size of the codebook.
\section{Conclusion}
In this paper, we proposed an optimization technique for the design of optimum transmit/receive waveforms in the discrete-time domain for FBMC systems, using hexagonal time-frequency lattices and operating over time and frequency dispersive channels. These waveforms
are obtained by using a new approach, known as POPS, for the maximization of the SINR or the SIR at the receiver.\\
We have shown that the optimal SIR obtained for FBMC with hexagonal lattices, outperforms the one obtained with rectangular lattices, especially for highly dispersive channels. Also, we have compared the performances of FBMC systems for both rectangular
and hexagonal lattices over time and frequency dispersive channels and demonstrated that the POPS-FBMC, with hexagonal lattices, is more efficient in the presence of frequency synchronisation errors and is more sensitive in the presence of time synchronisation errors. The latter characteristic, brought in part by the hexagonal nature of the time-frequency lattice, is not definitive and could be alleviated, and even inverted, using a non-optimal distribution of the Doppler spread and the time delay spread with respect to subcarrier frequency spacing and symbol duration. On the other hand, we have proved the dramatic increase in robustness brought by both systems relative to conventional OFDM, in the presence of time and frequency synchronisation errors. A spectral study was done and showed that the POPS-FBMC with hexagonal lattices offers a $10\hspace{0.1cm}\mbox{dB}$ (respectively, $70\hspace{0.1cm}\mbox{dB}$) reduction in OOB emissions with respect to FBMC with rectangular lattices (respectively, conventional OFDM). As potential future research works, we plan to consider different durations of the transmit and receive waveforms for POPS-FBMC, with hexagonal lattices, and extend our approach to continuous-time waveforms. We also intend to work on multipulse waveform design and waveform optimization for partial equalization, whereby interference from neighboring symbols, in time or/and in frequency, is tolerated thanks to a simplified equalization at the receiver.


\bibliographystyle{unsrt}
\bibliography{bibliography_hexagonal}

\end{document}